\documentclass[twocolumn]{aastex61} 

\usepackage{csvsimple}
\usepackage{multirow,bigdelim}
\usepackage{amsmath}

\newcommand{\msun}{\mbox{$M_\odot$}}

\newcommand{\mearth}{M$_\oplus$}

\newcommand{\ms}{\mbox{m s$^{-1}$}}

\usepackage{colortbl}
\definecolor{tablegray}{rgb}{0.89, 0.89, 0.89}

\begin{document}

\title{Stacked Periodograms as a Probe of Exoplanetary Populations}
\correspondingauthor{Samuel H. C. Cabot}

\email{sam.cabot@yale.edu}

\author{Samuel H. C. Cabot}
\affil{Yale University, 52 Hillhouse, New Haven, CT 06511, USA}

\author{Gregory Laughlin}
\affil{Yale University, 52 Hillhouse, New Haven, CT 06511, USA}

\begin{abstract}

Ongoing, extreme-precision Doppler radial velocity surveys seek planets with masses less than several \mearth; population-level studies to determine the distribution of planetary masses, however, remain difficult due to the required observational time investment, as well as challenges associated with robustly detecting the lowest mass planets. We outline a novel approach that leverages extensive, existing RV datasets to constrain masses of exoplanet populations: stacking periodograms of RV timeseries across many targets. We show that an exoplanet population may be statistically identifiable in the stacked periodogram, even when individual planets do not pass the threshold of detection. We discuss analytical, statistical properties of the stacked periodogram, perform simulations to demonstrate the efficacy of the method, and investigate the influence of semi-structured window functions and stellar activity. Analysis of the Lick-Carnegie Exoplanet Survey data set reveals a marginally significant ($1.6\sigma$) signal consistent with a population of exoplanets occupying $3-7$ day periods with typical $K$ between $1.6-5.1$ \ms. More detailed investigation of signals associated with stellar activity and yearly systematics may be necessary to confirm this result or detect other underlying Keplerian contributions. 

\end{abstract}


\section{Introduction} \label{sec:intro}

Transit surveys such as {\it Kepler} have densely populated much of the orbital parameter space with exoplanet detections \citep{Fressin2013}. The aggregate of photometric detections, however, is hindered by the lack of a correspondingly detailed corpus of planetary mass measurements. 
Important insights will likely follow in rapid succession if the exoplanet mass distribution is determined to an accuracy that is comparable to that enjoyed by the exoplanetary radii. 
 
The highest-yielding efforts to collect planetary masses (or specifically, $m\sin i$'s) are the Doppler radial velocity (RV) surveys \citep[e.g.][]{Howard2010}, which are shifting increasingly toward the detection and characterization of low-mass planets \citep{Fischer2016}. For decades, instrumental precision and limited observational cadence and baseline were the primary obstacles to planet detection; however, with the maturation of the RV technique, stellar activity has become the limiting factor for current-generation spectrographs \citep{PePe2013, Fischer2016, Jurgenson2016}. While there is a large-scale effort to model and remove the stellar activity contribution \citep{Dumusque2017}, the methods themselves have variable efficacy, and RV-based planet detections still require heavy observational time investments. These considerations pose a significant hurdle to population-level studies.

Here, we outline and explore the prospects for a new approach to constrain the population of low-mass planets orbiting nearby FGKM stars --- the stacked periodogram of residuals from RV surveys. While surveys generally seek statistically significant periodic signals in the time-series of individual stars, the residuals retain signatures of undetected planets, and their aggregate Fourier-space power is revealed in the stacked\footnote{{In this context, `stacked' means taking an average over multiple periodograms. This is not the same definition as used by \citet{Mortier2017}, who use `stacked' to describe the temporal evolution of a periodogram as more exposures are collected.}} periodogram. For a population of low-mass planets, the combined enhancement in power is of greater statistical significance than the signature of individual planets in their respective periodograms. As a case study, we analyze archived data published by the Lick-Carnegie Exoplanet Survey Team (LCES) \citep{Butler2017}, using RVs from this data set that have been corrected for systematics by \citet{Tal-Or2019}. 

The outline of this article is as follows. In \S\ref{sec:one}, we discuss the stacked periodogram from a theoretical standpoint, and delineate expectations for the features that correspond to low-mass planets with a given $K$ and $P$ distribution. Our treatment of the LCES data is discussed in \S\ref{sec:proc}. In \S\ref{sec:lcestheory}, we discuss  expectations for stacked periodograms that account for the specific characteristics of the LCES dataset (cadence, baseline, and uncertainties). The stacked periodogram of LCES residuals is analyzed in \S\ref{sec:res}, along with a characterization of the power distribution. Specifically, we model the contribution of stellar activity, and we determine the range of $K$-distributions that are consistent with the observed power. {\S\ref{sec:res} also provides a comparison between our methodology and periodogram stacking for benthic $\delta^{18}$O record analysis which has been used to establish the influence of Earth's orbital and precessional dynamics upon its climatic variations \citep{Lisiecki2005, Caminha-Maciel2019}}. Our conclusions, including an outlook for the application of the technique to extant and planned surveys are summarized in \S\ref{sec:con}.

{
\section{Properties of Stacked Periodograms} \label{sec:one}
}

{In order to establish baseline expectations for the stacked periodogram, we develop a simple analytical framework (Appendix~\ref{sec:appa}) that makes three assumptions: i.i.d. measurement perturbations and corresponding uncertainties; Keplerian signals without nuisance (e.g. stellar activity) signals; and that the timeseries have sufficiently dense coverage. We then compare it to the stacked periodogram of a simulated dataset. These ideal conditions are generally not upheld in real RV datasets, but provide a best-case-scenario testbed for the stacked periodogram, and whether it warrants application to archival datasets.}

{
\subsection{Comparison to Simulated Data}
}

\begin{figure*} 
\includegraphics[width=\linewidth]{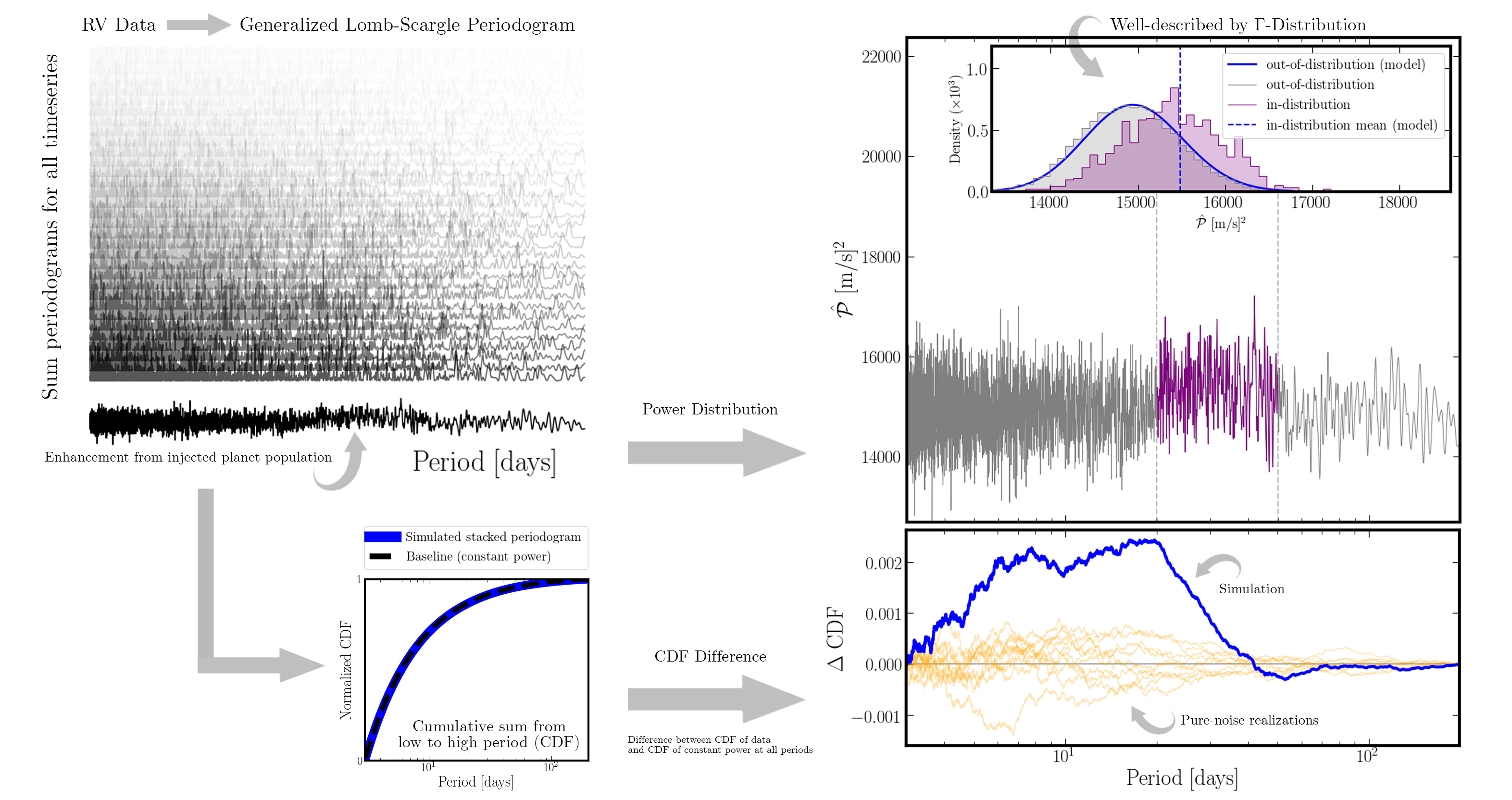}
\caption{Procedure for analyzing the stacked periodogram. The simulated periodograms (top left) involve 200 timestamps per 700 stars, each hosting a single planet in a circular orbit. The Keplerian signal is not statistically significant in most individual timeseries. However, the stacked periodogram (gray curve, top right), shows an enhancement in power (purple region) owing to the synthetic population of planets. The power in the region of interest ($20 - 50$ days, denoted `in-distribution') is visibly elevated compared to the remainder of the periodogram (denoted `out-of-distribution'). The distribution of power in the two regions is shown in the inset panel. The blue curve denotes the theoretical $\Gamma$-distribution describing `out-of-distribution' values, based on the simulation properties. The dashed blue line is the theoretical average enhancement in power from the Keplerian signals. The periodogram is unnormalized, hence the units of (\ms)$^2$. The CDF of the periodogram (bottom left) is normalized between zero and unity (blue curve). The CDF of a periodogram with equal power at all frequencies is also shown (labeled `Baseline', black dashed curve); although the two curves are essentially indistinguishable. The Keplerian contribution is apparent in the difference between the two CDFs (bottom right, blue solid curve). Noise realizations excluding the Keplerian contribution are shown in light orange to provide a sense of the significance of the planet population signal.}
\label{fig:sim1}
\end{figure*}

Consider the RV contribution of planets on circular orbits with
random orbital frequencies $\Delta\omega = {[1/50, 1/20]}$ day$^{-1}$ and with semi-amplitude $K = 1.5 \,\ms$. We simulated RV measurements for {$N_{\rm targets} = 700$} distinct systems, each hosting one planet, with Gaussian noise of amplitude $\sigma = 4.5$ \ms. We drew $N_{\rm obs} = 200$ timestamps uniformly at random from a $T = 5000$ day baseline for each system. Unnormalized periodograms were computed for each system across identical frequency grids, and subsequently co-added. Our analytical approximation (Appendix~\ref{sec:appa}) indicates the typical S/N of the periodogram in the frequency interval should be

\begin{equation} \label{eqn:snrmain}
    {\rm S/N} \approx \frac{2\pi\sqrt{N_{\rm targets}}N_{\rm obs}K^2}{4(\sigma^2 + K^2/2) T \Delta \omega} \approx 0.93
\end{equation}
per bin, up to a constant factor $\gamma \lesssim 1$ that approximates the leakage of power into other frequencies, as well as the noise introduced through the window function $W(t)$. While small, this S/N level is discernible over the $[1/50, 1/20]$ day$^{-1}$ frequency range as shown in Figure~\ref{fig:sim1}. The simulation agrees reasonably with the above prediction, as the average S/N in the frequency interval is 0.87. The planet's orbital period did not correspond to the periodogram maximum in most individual systems, and therefore was of insufficient strength to prompt a detection; however, the signal of the population is visibly noticeable in the stacked periodogram. In additional simulations, we found the S/N more closely agrees with the analytic prediction with higher cadence ($N_{\rm obs} \gtrsim 500)$ and measurement precision ($K/\sigma \gtrsim 1$), and that the region outside the enhancement agrees closely with its theoretical $\Gamma$-distribution. {While Equation~\ref{eqn:snrmain} works well in the dense, high-cadence limit, it is less accurate under conditions more typical of RV datasets.}

\subsection{The Periodogram's CDF}

It is important to gauge the significance of the enhancement in Figure~\ref{fig:sim1}. Equation~\ref{eqn:snrmain} does not return a detection significance of the entire feature, and requires {\it a priori} knowledge of the frequency range. As an alternative, we analyze the cumulative distribution function (CDF) of the periodogram, defined over the range [$\omega_{\rm min}, \omega_{\rm max}$] as

\begin{equation}
    {\rm CDF}(\omega) \equiv {\int_{\omega_{\rm max}}^{\omega}{\hat{\mathcal{P}}}d\omega } \Big/
    {\int_{\omega_{\rm max}}^{\omega_{\rm min}}{\hat{\mathcal{P}}}d\omega}.
\end{equation}
Next define the difference CDF as

\begin{equation}
    \Delta {\rm CDF}(\omega) \equiv \Big(\frac{\omega - \omega_{\rm max}}{\omega_{\rm min} - \omega_{\rm max}}\Big) - {\rm CDF}(\omega).
\end{equation}
The difference CDF is the fraction of total power in a periodogram when integrated up to $\omega_{\rm min}$, compared to the fraction of power expected if the periodogram is flat. In Figure~\ref{fig:sim1}, there is a high-confidence enhancement in the difference CDF at 20 days, which is due to the Keplerian component extending to 50 days. The enhancement at 20 days lies $>7\sigma$ from the baseline, where $\sigma$ is the standard deviation of $\Delta$CDF values from several noise realizations (the same simulation with $K = 0$ \ms). In the noise realizations, most of the variance in $\Delta$CDF is at shorter periods because we defined $\Delta$CDF as a function of frequency. 

Our analysis mimics some aspects of a two-sample Kolmogorov-Smirnov (KS) test --- however, the CDF as defined above concerns the power distribution across frequencies, as opposed to the probability of drawing a value of $\hat{\mathcal{P}}$. The KS test itself is unsuitable for this problem since, generally, the null-hypothesis distributions for $\hat{\mathcal{P}}$ lack a simple, closed form. The distributions are complicated by several observational and astrophysical processes discussed in the next sections. Further, the KS statistic reflects the supremum of the CDF separation, and neglects information regarding the full frequency distribution of Keplerian signals. 

\section{Data and Preprocessing} \label{sec:proc}

The LCES survey contains RVs collected over 20 years of monitoring nearby FGKM stars, and represents one of the more extensive RV datasets published to date. Specific details of the survey are presented by \citet{Butler2017}. In a later study of the dataset, \citet{Tal-Or2019} identified an instrumental offset introduced in 2004 as well as a long-term drift and an intra-night drift. Our analysis here adopts their published, corrected RVs.\footnote{https://cdsarc.unistra.fr/viz-bin/cat/J/MNRAS/484/L8}, and we restrict consideration to the 701 targets having both $N_{\rm obs} > 20$ RVs and $T > 1000$ day baselines. \citet{Butler2017} discuss previous planet claims and new planet candidates, but they did not publish the parameters from full Keplerian fits corresponding to all of the planets in the data. We thus carried out the following procedure to obtain RV residuals:
\begin{enumerate}
    \item Cross-referenced LCES targets with confirmed planet hosts in the NASA Exoplanet Archive \citep{exoarchive}.
    \item Retrieved recent entries for orbital parameters, for each planet in the system. In this step, it was ensured that the argument and the time of periastron were both from the same entry. If time of periastron was unavailable, it was calculated from time of conjunction/transit.
    \item Used \texttt{scipy.optimize} to fit a full Keplerian model to the data, setting initial guesses to the retrieved parameters. The model included a systemic offset, a jitter term (added in quadrature with measurement uncertainties), a linear trend term, and a quadratic trend term, regardless of whether the target hosts a confirmed planetary companion. Several intermediate fits with restricted parameters were performed to help reach convergence. The final fit always had all parameters free.
    \item If the residual RVs contained $>5\sigma$ outliers, these data points were removed and the Keplerian model was re-fit. This step removed $0.15\%$ of the RVs analyzed, and prevented outliers from biasing the stacked periodogram.
    \item Individual measurement uncertainties were changed to the quadrature sum of their original values and their respective timeseries' best-fit jitter value. Timestamps, residual RVs, and uncertainties were saved.
\end{enumerate}
We used the \texttt{radvel} Keplerian model \citep{Fulton2017, Fulton2018} throughout this analysis. We acknowledge that the linear and quadratic trend terms may not be statistically warranted in every fit, under a Bayesian analysis. However we included them to make the residual analysis more consistent, and to guarantee that gradual systematic trends and contributions from long-period companions were removed. The masses of LCES stars were obtained from the following surveys: \citet{Brewer2016}, \citet{Valenti2005}, and \citet{Schweitzer2019}. About $200$ stars lack mass measurements in these three studies. We assigned them each 1 \msun\ (the median mass of stars with measurements is 1.01 \msun). 

\section{Theoretical Expectations for the LCES Residuals}  \label{sec:lcestheory}

We performed additional simulations involving the identical timestamps of LCES data, and a multi-step analysis designed to replicate the generation of residual RVs in the survey. While still idealized, this test provides a more accurate gauge of expected features in the stacked periodogram, and the physical processes to which they correspond. {For the remainder of the study, we construct individual periodograms in the uncertainty-normalized, $\chi^2$ framework \citep{VanderPlas2018} in order to account for heteroscedastic RV measurements.}

\subsection{Simulating a Population of Small Planets}

\begin{figure} 
\includegraphics[width=\linewidth]{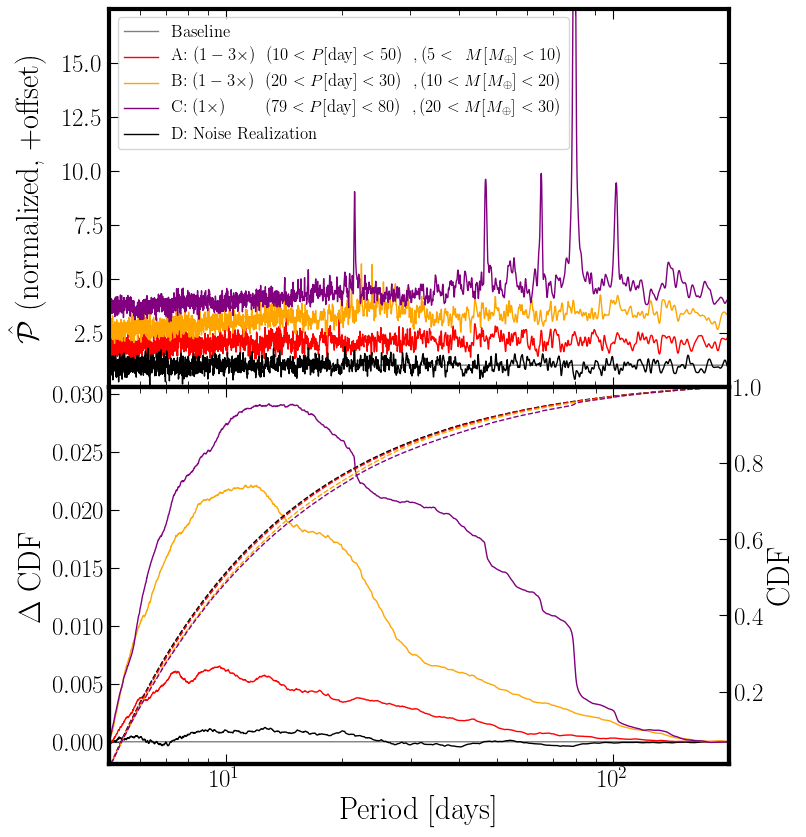}
\caption{{\it Top Panel}: Stacked periodograms for simulated RV timeseries with different injected planet populations, specified in the figure legend. The timestamps and uncertainties of the simulated RV timeseries are identical to those of the LCES dataset. Several peaks in the periodogram of simulation (C) arise from aliasing (e.g. as one-year peaks in the window function's periodogram). {\it Bottom Panel}: CDFs of the stacked periodograms shown by dotted lines. The subtle variations in the CDFs are revealed in the difference CDFs, obtained by subtracting the original CDF from another CDF that corresponds to a periodogram with constant power. Power enhancements are visible in the period ranges of the injected planets.}
\label{fig:sim2}
\end{figure}

First, synthetic RVs were generated assuming a small number of low-mass planets in each system. We added white noise to individual RVs with amplitudes matching the residual uncertainties. In the first simulation (A), we drew between $1-3$ planets with mass $5-10$ \mearth\ at periods between $10-50$ days. These parameters were drawn uniformly at random. The low-mass planets were ``short-period" per \citet{Kipping2013}, and we drew eccentricities accordingly as $e \sim \beta(a=0.697, b=3.27)$. Inclination $i$ was drawn from a sine distribution. We drew a reference time and argument of periastron from uniform distributions, $t_0 \sim \mathcal{U}(0, P)$ and $\omega \sim \mathcal{U}(0, 2\pi)$. It is emphasized that at this stage the adopted distributions are for demonstrating the efficacy of the method, and are not meant to represent actual exoplanet populations. The individual periodograms are dimensionless since residuals were first divided by their uncertainties (Appendix~\ref{sec:appa}), and the value of a single periodogram at a given frequency bin is distributed $\sim{\rm Exp}(\lambda=1)$. We simulated three other planet distributions, including (B) $1-3$ planets with mass $10-20$ \mearth\ at periods between $20-30$ days, (C) $1$ planet with mass $20-30$ \mearth\ at periods between $79-80$ days, and (D) a noise realization with no injected planets. The results of these tests are shown in Figure~\ref{fig:sim2}, and confirm that the quality and quantity of the LCES dataset is sufficient to resolve these planet populations (in the absence of nuisance signals, which we discuss below). 

\subsection{Simulating Stellar Activity}

Stellar activity is not perfectly coherent on long timescales, but often imparts a sufficiently strong periodic component in RVs to be detected in periodograms. At present, there is no accurate, analytic correspondence between activity-induced RVs and known indicators of activity, such as spectral features or photometry. Part of the challenge is that activity encompasses several physical processes that affect RVs on various timescales and at different amplitudes \citep{Fischer2016}. Nevertheless, we attempt investigate its effect on the stacked periodogram as follows. As in the previous simulation, we generated RVs by taking random draws from Gaussian distributions, but neglecting Keplerian contributions. The draws were scaled by the measurement uncertainties of the LCES residuals. For each timeseries, we sampled a Gaussian Process (GP) at the LCES timestamps to simulate spots and faculae on the rotating stellar surface \citep{Haywood2014, Angus2018}. We opted for the computationally efficient quasiperiodic GP kernel offered in \texttt{celerite} \citep{Foreman-Mackey2017}:

\begin{equation}
    K_{ij} = \frac{B}{2+C}e^{-|t_i-t_j|/L}\Big[\cos{\frac{2\pi|t_i-t_j|}{P_{\rm GP}}} + (1 + C)\Big].
    \label{eqn:k2}
\end{equation}
Hyperparameters $\phi = \{B, C, L, P_{\rm GP}\}$ are the magnitude of covariance, weighting of the periodic component, decay timescale, and rotation period, respectively. We fixed $C$ to a small positive value. For each timeseries, we drew $P_{\rm GP}\sim\mathcal{U}(5, 40)$ days, $L\sim\mathcal{U}(P_{\rm GP}, 3P_{\rm GP})$ days, and $B\sim\mathcal{U}(0, 9)$ (\ms)$^2$. Another GP was sampled and added to the RVs to simulate magnetic cycles. Its hyperparameters were drawn from the same distributions except $P_{\rm GP}\sim\mathcal{U}(2000, 5000)$ days. A best-fit quadratic polynomial was finally subtracted from the synthetic residuals to mimic the preprocessing of the LCES data. 

\begin{figure*}
\includegraphics[height=0.43\linewidth]{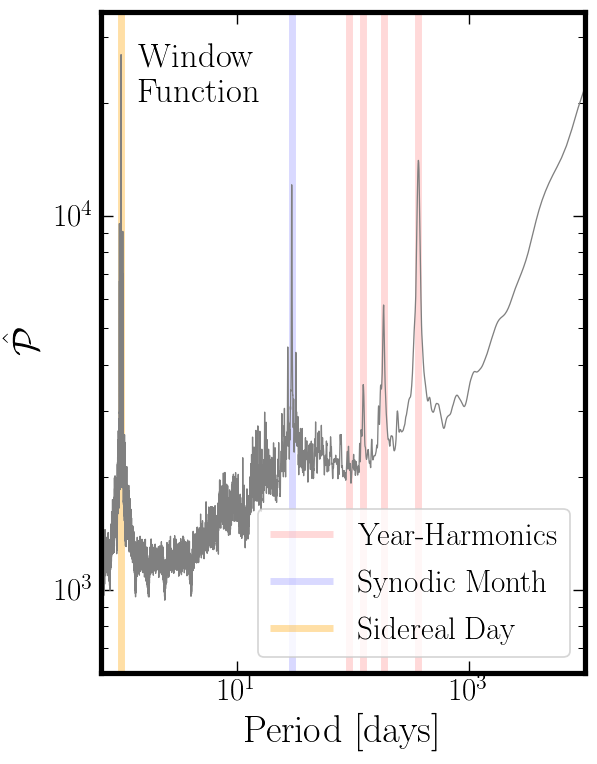} \includegraphics[height=0.43\linewidth]{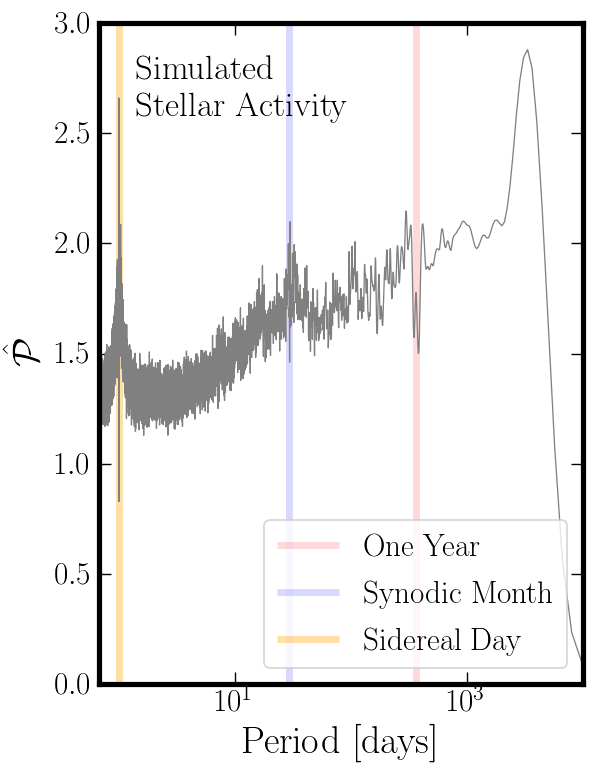} \includegraphics[height=0.43\linewidth]{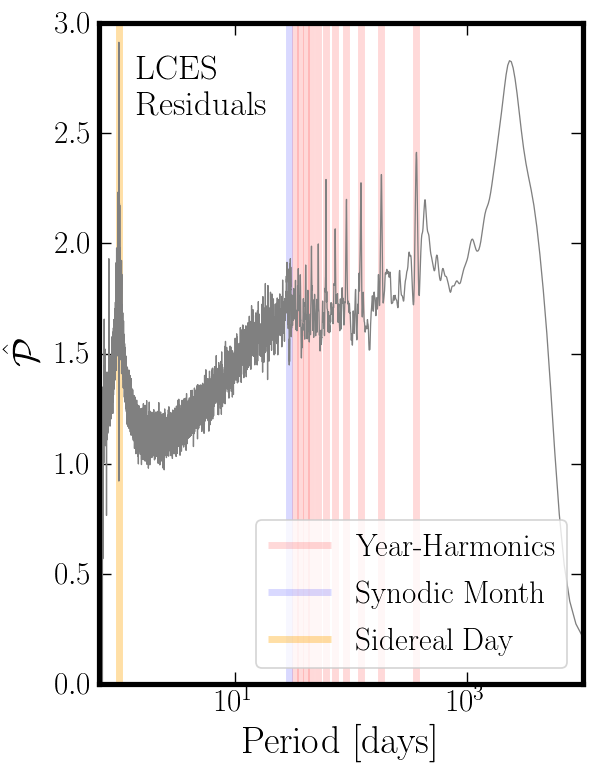} 
\caption{Comparison of stacked periodograms. {\it Left Panel}: Stack of periodograms of window functions for all LCES timeseries analyzed. The observing cadences generate a number of features, some of which are highlighted and labeled in this figure. Unlike the other panels, the window function has a logarithmic y-axis scale. {\it Middle Panel}: Stacked periodogram of simulated residuals, injected with quasiperiodic RV variations characteristic of stellar activity. Highlighted periods correspond to the strongest features in the window function's periodogram. {\it Right Panel}: Stacked periodogram of the real LCES residuals. Some of the prominent features match those in the window function's periodogram. However, several additional features are visible, such as higher harmonics ($1/4, 1/5, 1/6...$) of one year.}
\label{fig:single_per}
\end{figure*}

The stacked periodogram (Figure~\ref{fig:single_per}, {\it Middle Panel}) exhibits a gradual slope in power, similar to the window function's stacked periodogram (Figure~\ref{fig:single_per}, {\it Left Panel}). The large peak at a $P \sim$ a few thousand days is only apparent when we include the long-period, magnetic cycle GP. Periods past $\gtrsim 10,000$ days exceed the typical baselines of the timeseries, and power is removed along with the quadratic polynomial. Both panels highlight several of the strongest features in the window function's periodogram --- which is also where we expect aliasing. Magnetic cycles visibly alias with one year, yielding peaks on either side of the gray vertical line (Figure~\ref{fig:single_per}, {\it Middle Panel}). All of the injected signals alias with one sidereal day. Aliasing of a frequency $f$ with a sampling frequency $f_s$ enhances power at $f\pm nf_s$ for integers $n$, but not at $f_s$; hence the lack of power at exactly one year (and at one sidereal day, which is not immediately discernible in Figure~\ref{fig:single_per}, {\it Middle Panel}). Aliasing is expected near $29.5$ days (the synodic month), but its power is not distinguishable from that of the injected stellar rotation signals. 

\setlength{\tabcolsep}{12pt}

\section{Signals in the LCES Residuals} \label{sec:res}

The stacked periodogram of LCES residuals (Section~\ref{sec:proc}) is shown in Figure~\ref{fig:single_per} ({\it Right Panel}). We proceed to analyze the power distribution and identify contributing sources. The most conspicuous features in the stacked periodogram are: (1) a peak at $\sim 2200$ days, likely associated with stellar activity cycles. The median baseline of the LCES observations is about 5300 days, which is sufficient to capture up to several full cycles. The quadratic trend removal eliminates power at periods past several thousand days; (2) a feature at one sidereal day, broadened by aliases of  other intrinsic sources of power; (3) a gradual increase in power extending from about $4-2000$ days. A similar trend is apparent in the window function's periodogram; however, certain sections of the periodogram may be influenced by rotationally modulated stellar activity or by low-mass planets; and (4) sharp peaks at one year and many harmonics of one year. \citet{Rosenthal2021} investigated archival RVs significantly overlapping with the LCES dataset considered here, with a focus on detecting individual planets. They identified several RV timeseries that correlated with systematics (e.g. PSF parameters), which resulted in power at one year or its harmonics. Indeed, since aliasing would not produce power at these exact frequencies, the features are most likely associated with systematics. The one year feature is broadened, however, due to aliasing with long period power.

\subsection{Isolating the Stellar Activity Contribution}

Our simple stellar activity simulation yields a periodogram strikingly similar to that of the LCES residuals. We refrain from performing a formal fit, given that GPs are a simplified, {\it ad hoc} model of stellar activity, and that a fit would require detailed modeling of yearly systematics, which is beyond the scope of this paper. However, the similarity indicates it is challenging to directly infer the presence of a Keplerian component. We attempt to separate the Keplerian and activity contributions by examining periodogram power dependence on the $\log R'_{hk}$ activity indicator, which measures the chromospheric emission in H and K Ca lines. The indicator $\log R'_{hk}$ should track power in the rotation period regime (approximately $5-40$ days) and the magnetic cycle regime (approximately $2000-5000$ days), whereas a Keplerian population component should be independent of the stellar activity level, neglecting correlations between stellar properties and planet types, formation, and multiplicity.

\begin{figure*} 
\includegraphics[width=\linewidth]{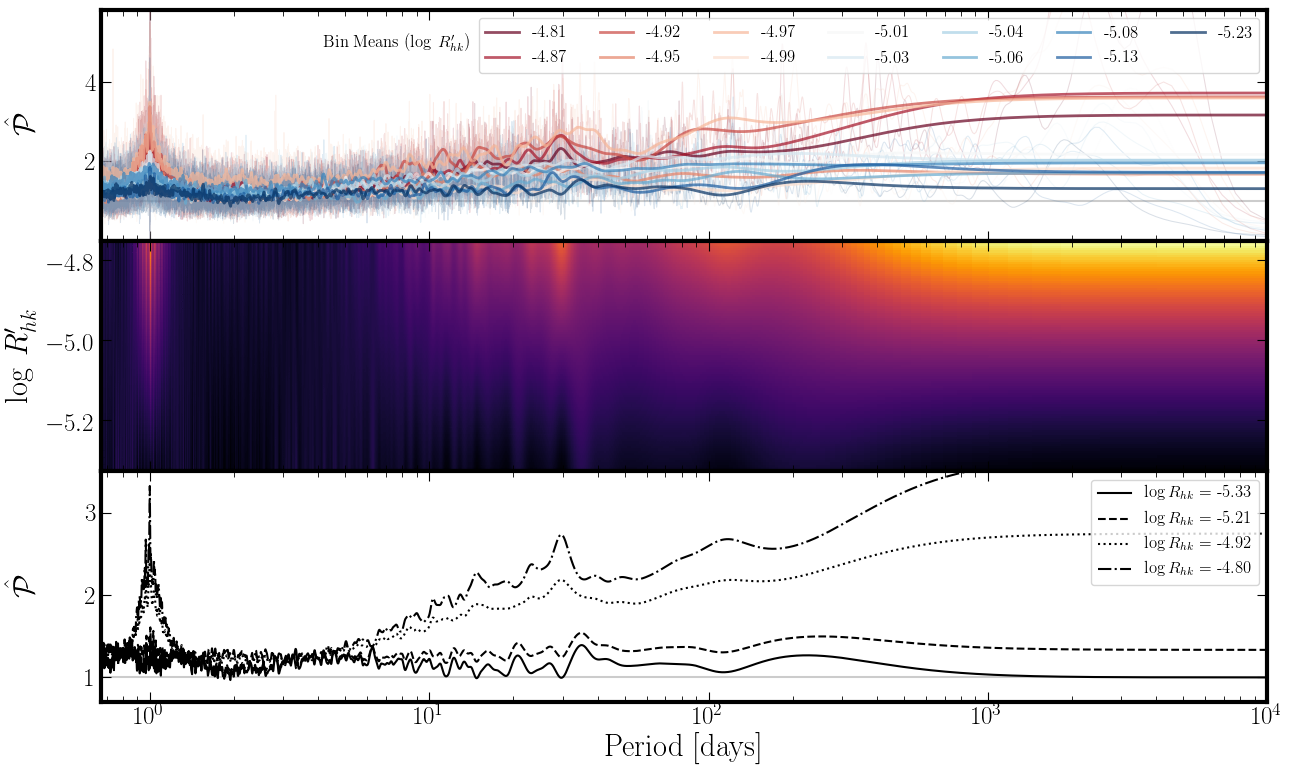}
\caption{Dependence of stacked periodogram power on $\log R'_{hk}$. {\it Top Panel}: Subsets of the LCES periodograms were stacked based on their $\log R'_{hk}$ (bin means reported in the figure legend). The stacks were then smoothed with a Gaussian kernel. The color-code ranges from blue (lowest $\log R'_{hk}$) to red (highest $\log R'_{hk}$), and the un-smoothed stacked periodograms are shown in gray at low opacity. The trend appears consistent with expectations -- for example, long-period power associated with magnetic activity cycles is greater for larger $\log R'_{hk}$. {\it Middle Panel}: An exponential function was fit to the power in every frequency bin. The heatmap shows the interpolated power for a range of $\log R'_{hk}$ values. {\it Bottom Panel}: The exponential functions were subsequently evaluated at several different $\log R'_{hk}$ values. This panel shows the resulting periodograms. Of particular interest is the $\log R'_{hk} = -5.33$ evaluation which is lowest $\log R'_{hk}$ value of the LCES target list (i.e. the most quiescent star).}
\label{fig:rhk}
\end{figure*}

Values of $\log R'_{hk}$ were obtained from \citet{Brewer2016} for 361 stars in the LCES sample. The spectral S-index, which measures both chromospheric and photospheric contributions to the H and K Ca emission-line cores, was also considered as an activity-proxy; however, we opted to use $\log R'_{hk}$ since it is known to be better correlated with stellar activity levels (i.e. presence of spots and plages) \citep{Noyes1984}. We found some degeneracy between $\log R'_{hk}$ and S-index in our sample. For example, targets with S-index near 0.15 spanned $\log R'_{hk}$ from approximately $-5.1$ to $-4.8$. Targets with $\log R'_{hk} > -4.75$ exhibited extremely poor correlation with S-value and surprisingly low-power in their respective periodograms. They also lie considerably far from the bulk of the $\log R'_{hk}$, with median $-4.99$ and standard deviation $0.19$. We excluded them in the following analysis, leaving 321 targets with median $\log R'_{hk}$ of $-5.01$ and standard deviation $0.11$. The targets were sorted by their $\log R'_{hk}$ values and grouped into equal-sized bins (25 targets per bin, except for the highest $\log R'_{hk}$ bin containing 21 targets). The periodograms in each bin were stacked and subsequently convolved with a Gaussian smoothing kernel of standard deviation $0.0015$ Hz (101 elements in the frequency grid), the result depicted in Figure~\ref{fig:rhk} ({\it Top Panel}). Significant correlation is apparent between $\log R'_{hk}$ and power at most periods, which is expected since stars of higher $\log R'_{hk}$ have larger spot coverage and greater amplitude of activity-based RV variations. At each frequency bin $n$, we fit this relationship with an exponential function: $\log{\mathcal{\hat{P}}} = a_n\log R'_{hk} + b_n$. The fit is shown in Figure~\ref{fig:rhk} ({\it Middle Panel}), along with several example draws from the fit ({\it Bottom Panel}). 

\subsection{Constraints on Planet Populations}

\begin{figure} 
\includegraphics[width=\linewidth]{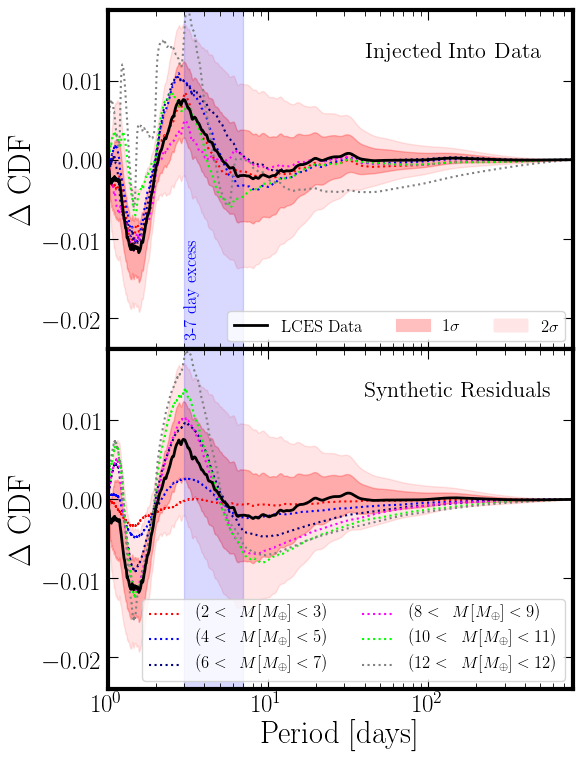}
\caption{Constraints on Keplerian contributions in the stacked periodogram, which was evaluated at the lowest $\log R'_{hk}$ in the sample. The black curve depicts the observed $\Delta$CDF. A bootstrap analysis yielded uncertainties on $\Delta$CDF, shown as light red shaded regions. The blue vertical band highlights an excess of power between 3 and 7 days. Keplerian signals were injected into the LCES residuals, with masses specified in the legend, and multiplicity between 1 and 3. The results are shown in the top panel as dotted color lines. As another test, synthetic residuals were generated assuming white noise with the timestamps and uncertainties of the LCES dataset. Random planet distributions were again injected, and the results are shown in the bottom panel. The best agreement with the observed $\Delta$CDF corresponds to masses of $6-7$ $M_{\oplus}$ (dark blue dotted line).}
\label{fig:planetpop}
\end{figure}

An undetected planet population orbiting the LCES stellar sample should have a consistent power signature across $\log R'_{hk}$, whereas power from stellar activity depends strongly on $\log R'_{hk}$. The exponential fit above was evaluated at its lower limit ($\log R'_{hk} = -5.33$). Apparent in Figure~\ref{fig:rhk} ({\it Bottom Panel}), the power in this periodogram largely lies at unity, with the exception of some small scale features. While most of the features are very low amplitude, there is an excess of power between approximately 3 and 7 days, more clearly seen in the difference CDF (Figure~\ref{fig:planetpop}). The other major feature in $\Delta$CDF is due to aliasing near one sidereal day. Power remains in this region both due to intrinsic features, such as planets, and possibly due to imperfections when fitting the exponential form across $\log R'_{hk}$. {For comparison, we repeated the above analysis, but instead fit against S-index, and then H-index as the activity-proxy. S-index and H-index are both included in the LCES data release. When selecting an activity bin for a target, we used the median value of the proxy over all spectra of that target. For both proxies, we found qualitatively similar results as those in Figure~\ref{fig:rhk}: correlation between the activity-proxy and the power in the stacked periodograms, and an enhancement in the $3-7$ day regime when we evaluated at the lowest activity value in the sample. H-index had the least coherent trend, especially at long periods ($100-1000$ days). For S-index, when we evaluated at the lowest activity level, the periodogram had features in similar locations and of similar heights as in the $\log R'_{hk}$ case. Moving forward, we use results from our $\log R'_{hk}$ analysis.}

We performed a bootstrap analysis consisting of 50 runs, each repeating the analysis of the previous subsection (stacking periodograms in $\log R'_{hk}$ bins, fitting an exponential, and evaluating at $\log R'_{hk} = -5.33$), but with random sampling with replacement of targets. Then $\Delta$CDF was calculated for each run, and the standard deviation across each frequency was used as an estimate of uncertainty in the actual $\Delta$CDF. The uncertainty contours are plotted in Figure~\ref{fig:planetpop}. The power enhancement at $\sim 3$ days is of approximately $1.6\sigma$ significance, and is then accounted for by the slope from $3-7$ days. 

We explored a range of planet masses that would produce a similar signature, while restricting the population to periods between 3 and 7 days (sampled uniformly at random) and between 1 and 3 planets per system. As a first test, planets were injected into the real LCES dataset and the activity signal was again removed via the $\log R'_{hk}$ exponential fitting method above. Results are depicted in Figure~\ref{fig:planetpop} ({\it Top Panel}). The calculated $\Delta$CDFs exhibit high variance at low periods. In most cases, the additional planet burden yields a comparable or greater power enhancement at 3 days and subsequent slope up to 7 days, compared to the original data's $\Delta$CDF. The erratic behavior in $\Delta$CDF is probably due to spurious interactions the added Keplerians have with the window function and yearly systematics. Furthermore, $\Delta$CDF of the LCES data is fairly insensitive to added planet burdens of $\lesssim 6 M_{\oplus}$ per system, as they lie within the $1\sigma$ contours. Next, we simulated planet populations atop white noise, as done in Section~\ref{sec:lcestheory}. We find good agreement with the observed $\Delta$CDF --- in particular, we find $1-3$ planets with $6-7$ $M_{\oplus}$ per planet lies within the $1\sigma$ contours for nearly the entire period range. The largest deviation is near one day, a regime which is significantly complicated aliasing. 

{
\subsection{Alternative Computation of the Stacked Periodogram}
}

{
There is precedent for stacking Lomb-Scargle Periodograms of geophysical timeseries. \citet{Caminha-Maciel2019} developed the \texttt{LSTperiod} software, which they then applied to $\delta^{18}$O variations in sedimentary cores \citep{Lisiecki2005}. They recovered the obliquity (41 kyr) and precession (19 kyr and 23 kyr) Milankovitch cycles in the stacked periodogram at higher S/N compared to individual periodograms. As a foil to our analysis, we used \texttt{LSTperiod} software on the LCES residuals (\S\ref{sec:proc}).
\texttt{LSTperiod} accepts multiple, two-column (timestamp, value) timeseries. It then computes the periodogram of each timeseries and normalizes its integrated power to unity. Then, it combines the periodograms through multiplication (which produces the ``AND" periodogram) and addition (which produces the ``OR" periodogram). Power in the stacked periodogram is then normalized to $2\pi$. The ``OR" periodogram is most relevant to our study since a planet population would introduce power across a range of periods; the ``AND" periodogram is more appropriate in cases where the timeseries exhibit modes at identical frequencies.
}

\begin{figure} 
\includegraphics[width=\linewidth]{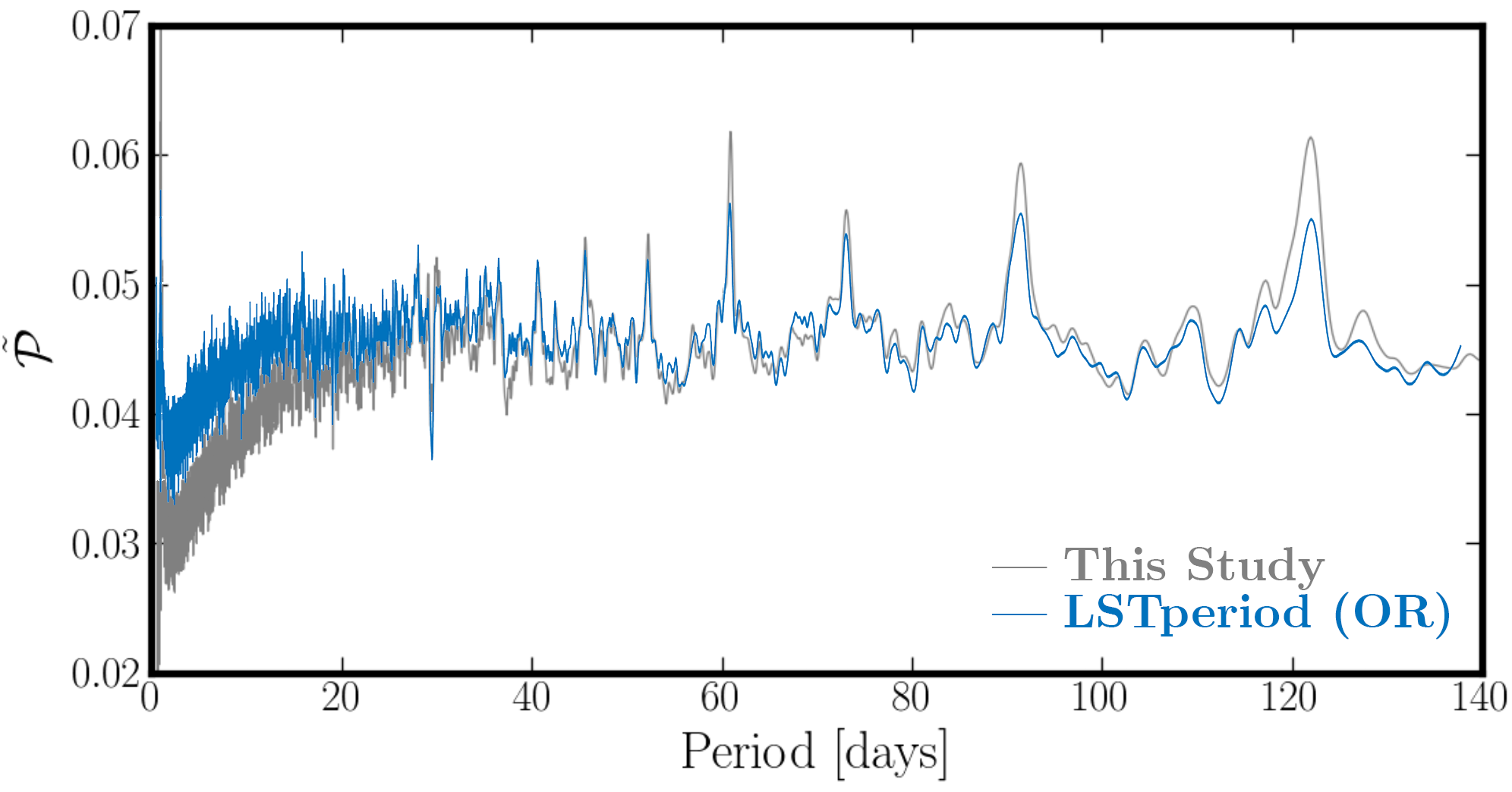}
\caption{{Comparison of our stacked periodogram (This Study) to the \texttt{LSTperiod} ``OR" periodogram \citep{Caminha-Maciel2019}. The periodograms were computed using the identical LCES residuals; however, our stacking approach makes use of individual, uncertainty-normalized periodograms, whereas \texttt{LSTperiod} does not use uncertainties, and instead normalizes individual periodograms to unity. To match the \texttt{LSTperiod} convention, we renormalized our stacked periodogram to integrate to $2\pi$ (power is denoted by $\tilde{\mathcal{P}}$).}}
\label{fig:lst}
\end{figure}

{
Our stacked periodogram is very similar to the \texttt{LSTperiod} ``OR" periodogram, as seen in Figure~\ref{fig:lst}. Both exhibit a steep rise in power up to $\sim 20$ days, and strong peaks at one-year harmonics. They also have similar fine-scale structure.
}

{
The main difference between the two is the offset at periods $\lesssim 40$ days, which is most likely due to the different treatment of individual timeseries. By its normalization, \texttt{LSTperiod} assigns the periodogram of each timeseries equal weight in the arithmetic mean. Since our approach uses the $\chi^2$ framework, periodograms of timeseries that have few data points and larger uncertainties receive lower weights. Both schemes ensure that a single, erratic timeseries will not dominate the combined spectrum.
}

\section{Discussion and Conclusions} \label{sec:con}

We have demonstrated the efficacy of stacked periodograms for revealing signatures of planet populations in RV datasets. Specifically, we focus on low-$K$ planets which are undetectable in their respective host's periodograms. The stacked periodogram adheres to relatively simple statistical properties. We performed several simulations of various planet populations and data quantity/quality to obtain performance expectations. As a case study, we examine one of the largest archival RV datasets \citep[LCES][]{Butler2017}, and assess the detectability of low-mass planet populations. While the measurement uncertainty, time baseline, and number of exposures and targets are sufficient for detecting populations of several $M_{\oplus}$ planets, complications arise from observational-related systematics, a semi-structured window function, and power associated with stellar rotation and magnetic activity cycles. Through an {\it ad hoc} functional fit between the activity indicator $\log R'_{hk}$ and Fourier power, we interpolate the stacked periodogram at the lowest activity level of all LCES targets. A marginally significant power excess is observed at $3-7$ days, which is consistent with between one and a few planets per system, each of several $M_{\oplus}$. While a limited region of parameter space is explored, we find particularly good agreement with a population of $1-3$ planets with $6-7$ $M_{\oplus}$ per planet, or approximately $1.6-5.1$ \ms\ of additional $K$ per star (which represent the $16^{\rm th}$ and $84^{\rm th}$ percentiles of the sum of $K$ per star injected, accounting for random eccentricity and inclination). 
We also emphasize the importance of understanding non-Keplerian sources of Fourier power, which often obscure planetary signals or are the source of false-positive detections in conventional planet searches. The dependence of the stacked periodogram on the stellar activity indicator $\log R'_{hk}$ is fully within expectations: strong correlation at $\gtrsim$ 1000 days due to magnetic cycles, moderate correlation between $5-100$ days due to stellar rotation, and strong correlation within peaks of the window function (e.g. one sidereal day and one synodic month). 

Over 350 {\it Kepler} multi-planet systems are known \citep{Rowe2014, Weiss2018}, which contain {more than} 900 planets. They exhibit a surprising uniformity in radius and mass \citep{Weiss2018, Millholland2017}, with typical masses less than 10 $M_{\oplus}$ (as measured from transit-timing-variations); many of these systems also host planets with orbital periods of $10$ days or less. Recently, \citet{Rosenthal2021} conducted a planet search within archival RVs that significantly overlap with the LCES dataset, plus the addition of a couple additional surveys. They recovered {43} planets with $M_p\sin i < $ 30 $M_{\oplus}$ and {20} planets with $M_p\sin i < $ 10 $M_{\oplus}$ across 719 target stars. Non-transiting multi-planet systems \citep[e.g.][]{Udry2019} are rarer, and only several are reported in the recent {\it California Legacy Survey} planet catalog \citep{Rosenthal2021}. The power excess identified in our LCES stacked periodogram is consistent with this population of planets --- however the feature is at low significance, and will require additional data to verify. Furthermore, the signature of the population of planets orbiting at $\gtrsim 10$ days is largely lost within contamination from stellar rotation.

Statistics for planets $R_{p} \lesssim 1.25R_\oplus$ are subject to diminishing completeness at $3-7$ day periods \citep{Fulton2017b}, and planets smaller than $0.5R_\oplus$ are nearly undetectable ($< 10\%$ pipeline completeness) \citep{Burke2015}. \citet{Hsu2018} revise-up occurrence rates ($f$) for planets smaller than the \citet{Fulton2017b} valley minimum ($R_{p} \lesssim 1.75R_\oplus$) --- up to $f \approx 0.2$ planets per star for $R_{p} < 1.75R_\oplus$ and periods between $5-10$ days, plus $f \approx 0.08$ for periods between $2.5-5$ days. The occurrence rate increases at smaller $R_{p}$ for the broader $0.5-80$ day period range, and becomes poorly constrained at $R_{p} < 1.0R_\oplus$. 

Could undetected small planets contribute an excess of $K$ in RV surveys? Considering uncertainty in the mass-radius relation \citep{Chen2017}, a $R_{p} = 2R_\oplus$ planet could reasonably have $M_p = 3-9 M_\oplus$, and a $R_{p} = 1R_\oplus$ planet $M_p = 0.6-1.5 M_\oplus$. Fe-rich planets like K2-229b \citep[$R_p = 1.165\pm0.066 R_\oplus$ and $M_p = 2.59\pm0.43 M_\oplus$,][]{Santerne2018} could also contribute to the necessary mass. For example, a planet with radius $1-1.25R_\oplus$ could have a mass between $2-5 M_\oplus$ \citep{Zeng2019}. Taking semi-amplitude $K \simeq 0.64 \, \ms (M_p\sin i/M_\oplus)(P/1{\rm \, day})^{-1/3}$ for circular orbits around Sun-like stars, a $3 M_\oplus$ planet on a 5-day orbit with median inclination $\sin i = 0.866$ has $K \approx 1$ \ms. If `peas-in-a-pod' configurations \citep{Weiss2018, Millholland2017} of such planets are common, then they may form a population of planets that is largely undetectable by {\it Kepler} and simultaneously responsible for the additional $K$ in our stacked-periodogram signal. 

The approach presented here may serve as a useful tactic for forthcoming RV surveys to gauge the baseline power across many targets in the absence of stellar activity. Indeed, it appears chromospherically-quiet stars may be most amenable for such population-level studies. The Kepler occurrence statistics indicate a considerable fraction of Earth to Neptune-sized planets at periods between 10 and 85 days \citep{Fressin2013}. The stacked RV analysis considered here shows robust and predictable scaling with the number of targets and uncertainty on RV measurements, but information is lost in this crucial period range since it overlaps with typical stellar rotation periods. Therefore, while bright (e.g. $V < 8$) stars are typical for many RV programs, exceptionally low-activity stars (e.g. $\log R'_{hk} \lesssim -5.3$, possibly at a cost of higher $V$) may better serve long-term strategies that involve stacking across multiple sources.

\acknowledgements

We acknowledge generous support from the Heising-Simons Foundation through Grant $\#$2021-2802 to Yale University. This research has made use of the NASA Exoplanet Archive, which is operated by the California Institute of Technology, under contract with the National Aeronautics and Space Administration under the Exoplanet Exploration Program.

\bibliographystyle{aasjournal} 
\bibliography{main}

\begin{thebibliography}{}
\expandafter\ifx\csname natexlab\endcsname\relax\def\natexlab#1{#1}\fi
\providecommand{\url}[1]{\href{#1}{#1}}
\providecommand{\dodoi}[1]{doi:~\href{http://doi.org/#1}{\nolinkurl{#1}}}
\providecommand{\doeprint}[1]{\href{http://ascl.net/#1}{\nolinkurl{http://ascl.net/#1}}}
\providecommand{\doarXiv}[1]{\href{https://arxiv.org/abs/#1}{\nolinkurl{https://arxiv.org/abs/#1}}}

\bibitem[{{Angus} {et~al.}(2018){Angus}, {Morton}, {Aigrain}, {Foreman-Mackey},
  \& {Rajpaul}}]{Angus2018}
{Angus}, R., {Morton}, T., {Aigrain}, S., {Foreman-Mackey}, D., \& {Rajpaul},
  V. 2018, \mnras, 474, 2094, \dodoi{10.1093/mnras/stx2109}

\bibitem[{{Baluev}(2008)}]{Baluev2008}
{Baluev}, R.~V. 2008, \mnras, 385, 1279,
  \dodoi{10.1111/j.1365-2966.2008.12689.x}

\bibitem[{{Brewer} {et~al.}(2016){Brewer}, {Fischer}, {Valenti}, \&
  {Piskunov}}]{Brewer2016}
{Brewer}, J.~M., {Fischer}, D.~A., {Valenti}, J.~A., \& {Piskunov}, N. 2016,
  \apjs, 225, 32, \dodoi{10.3847/0067-0049/225/2/32}

\bibitem[{{Burke} {et~al.}(2015){Burke}, {Christiansen}, {Mullally}, {Seader},
  {Huber}, {Rowe}, {Coughlin}, {Thompson}, {Catanzarite}, {Clarke}, {Morton},
  {Caldwell}, {Bryson}, {Haas}, {Batalha}, {Jenkins}, {Tenenbaum}, {Twicken},
  {Li}, {Quintana}, {Barclay}, {Henze}, {Borucki}, {Howell}, \&
  {Still}}]{Burke2015}
{Burke}, C.~J., {Christiansen}, J.~L., {Mullally}, F., {et~al.} 2015, \apj,
  809, 8, \dodoi{10.1088/0004-637X/809/1/8}

\bibitem[{{Butler} {et~al.}(2017){Butler}, {Vogt}, {Laughlin}, {Burt},
  {Rivera}, {Tuomi}, {Teske}, {Arriagada}, {Diaz}, {Holden}, \&
  {Keiser}}]{Butler2017}
{Butler}, R.~P., {Vogt}, S.~S., {Laughlin}, G., {et~al.} 2017, \aj, 153, 208,
  \dodoi{10.3847/1538-3881/aa66ca}

\bibitem[{Caminha-Maciel \& Ernesto(2019)}]{Caminha-Maciel2019}
Caminha-Maciel, G., \& Ernesto, M. 2019, Annals of Geophysics, 62,
  \dodoi{10.4401/ag-7923}

\bibitem[{{Chen} \& {Kipping}(2017)}]{Chen2017}
{Chen}, J., \& {Kipping}, D. 2017, \apj, 834, 17,
  \dodoi{10.3847/1538-4357/834/1/17}

\bibitem[{{Delisle} {et~al.}(2020){Delisle}, {Hara}, \&
  {S{\'e}gransan}}]{Delisle2020}
{Delisle}, J.~B., {Hara}, N., \& {S{\'e}gransan}, D. 2020, \aap, 635, A83,
  \dodoi{10.1051/0004-6361/201936905}

\bibitem[{{Dumusque} {et~al.}(2017){Dumusque}, {Borsa}, {Damasso}, {D{\'\i}az},
  {Gregory}, {Hara}, {Hatzes}, {Rajpaul}, {Tuomi}, {Aigrain},
  {Anglada-Escud{\'e}}, {Bonomo}, {Bou{\'e}}, {Dauvergne}, {Frustagli},
  {Giacobbe}, {Haywood}, {Jones}, {Laskar}, {Pinamonti}, {Poretti}, {Rainer},
  {S{\'e}gransan}, {Sozzetti}, \& {Udry}}]{Dumusque2017}
{Dumusque}, X., {Borsa}, F., {Damasso}, M., {et~al.} 2017, \aap, 598, A133,
  \dodoi{10.1051/0004-6361/201628671}

\bibitem[{{Fischer} {et~al.}(2016){Fischer}, {Anglada-Escude}, {Arriagada},
  {Baluev}, {Bean}, {Bouchy}, {Buchhave}, {Carroll}, {Chakraborty}, {Crepp},
  {Dawson}, {Diddams}, {Dumusque}, {Eastman}, {Endl}, {Figueira}, {Ford},
  {Foreman-Mackey}, {Fournier}, {F{\H{u}}r{\'e}sz}, {Gaudi}, {Gregory},
  {Grundahl}, {Hatzes}, {H{\'e}brard}, {Herrero}, {Hogg}, {Howard}, {Johnson},
  {Jorden}, {Jurgenson}, {Latham}, {Laughlin}, {Loredo}, {Lovis}, {Mahadevan},
  {McCracken}, {Pepe}, {Perez}, {Phillips}, {Plavchan}, {Prato}, {Quirrenbach},
  {Reiners}, {Robertson}, {Santos}, {Sawyer}, {Segransan}, {Sozzetti},
  {Steinmetz}, {Szentgyorgyi}, {Udry}, {Valenti}, {Wang}, {Wittenmyer}, \&
  {Wright}}]{Fischer2016}
{Fischer}, D.~A., {Anglada-Escude}, G., {Arriagada}, P., {et~al.} 2016, \pasp,
  128, 066001, \dodoi{10.1088/1538-3873/128/964/066001}

\bibitem[{{Foreman-Mackey} {et~al.}(2017){Foreman-Mackey}, {Agol},
  {Ambikasaran}, \& {Angus}}]{Foreman-Mackey2017}
{Foreman-Mackey}, D., {Agol}, E., {Ambikasaran}, S., \& {Angus}, R. 2017, \aj,
  154, 220, \dodoi{10.3847/1538-3881/aa9332}

\bibitem[{{Fressin} {et~al.}(2013){Fressin}, {Torres}, {Charbonneau}, {Bryson},
  {Christiansen}, {Dressing}, {Jenkins}, {Walkowicz}, \&
  {Batalha}}]{Fressin2013}
{Fressin}, F., {Torres}, G., {Charbonneau}, D., {et~al.} 2013, \apj, 766, 81,
  \dodoi{10.1088/0004-637X/766/2/81}

\bibitem[{Fulton \& Petigura(2017)}]{Fulton2017}
Fulton, B., \& Petigura, E. 2017, RadVel: Radial Velocity Fitting Toolkit,
  v0.9.1,  Zenodo, \dodoi{10.5281/zenodo.580821}

\bibitem[{{Fulton} {et~al.}(2018){Fulton}, {Petigura}, {Blunt}, \&
  {Sinukoff}}]{Fulton2018}
{Fulton}, B.~J., {Petigura}, E.~A., {Blunt}, S., \& {Sinukoff}, E. 2018, \pasp,
  130, 044504, \dodoi{10.1088/1538-3873/aaaaa8}

\bibitem[{{Fulton} {et~al.}(2017){Fulton}, {Petigura}, {Howard}, {Isaacson},
  {Marcy}, {Cargile}, {Hebb}, {Weiss}, {Johnson}, {Morton}, {Sinukoff},
  {Crossfield}, \& {Hirsch}}]{Fulton2017b}
{Fulton}, B.~J., {Petigura}, E.~A., {Howard}, A.~W., {et~al.} 2017, \aj, 154,
  109, \dodoi{10.3847/1538-3881/aa80eb}

\bibitem[{{Groth}(1975)}]{Groth1975}
{Groth}, E.~J. 1975, \apjs, 29, 285, \dodoi{10.1086/190343}

\bibitem[{{Haywood} {et~al.}(2014){Haywood}, {Collier Cameron}, {Queloz},
  {Barros}, {Deleuil}, {Fares}, {Gillon}, {Lanza}, {Lovis}, {Moutou}, {Pepe},
  {Pollacco}, {Santerne}, {S{\'e}gransan}, \& {Unruh}}]{Haywood2014}
{Haywood}, R.~D., {Collier Cameron}, A., {Queloz}, D., {et~al.} 2014, \mnras,
  443, 2517, \dodoi{10.1093/mnras/stu1320}

\bibitem[{{Howard} {et~al.}(2010){Howard}, {Marcy}, {Johnson}, {Fischer},
  {Wright}, {Isaacson}, {Valenti}, {Anderson}, {Lin}, \& {Ida}}]{Howard2010}
{Howard}, A.~W., {Marcy}, G.~W., {Johnson}, J.~A., {et~al.} 2010, Science, 330,
  653, \dodoi{10.1126/science.1194854}

\bibitem[{{Hsu} {et~al.}(2018){Hsu}, {Ford}, {Ragozzine}, \&
  {Morehead}}]{Hsu2018}
{Hsu}, D.~C., {Ford}, E.~B., {Ragozzine}, D., \& {Morehead}, R.~C. 2018, \aj,
  155, 205, \dodoi{10.3847/1538-3881/aab9a8}

\bibitem[{{Jurgenson} {et~al.}(2016){Jurgenson}, {Fischer}, {McCracken},
  {Sawyer}, {Szymkowiak}, {Davis}, {Muller}, \& {Santoro}}]{Jurgenson2016}
{Jurgenson}, C., {Fischer}, D., {McCracken}, T., {et~al.} 2016, Society of
  Photo-Optical Instrumentation Engineers (SPIE) Conference Series, Vol. 9908,
  {EXPRES: a next generation RV spectrograph in the search for earth-like
  worlds}, 99086T, \dodoi{10.1117/12.2233002}

\bibitem[{{Kipping}(2013)}]{Kipping2013}
{Kipping}, D.~M. 2013, \mnras, 434, L51, \dodoi{10.1093/mnrasl/slt075}

\bibitem[{{Koen}(1990)}]{Koen1990}
{Koen}, C. 1990, \apj, 348, 700, \dodoi{10.1086/168277}

\bibitem[{Lisiecki \& Raymo(2005)}]{Lisiecki2005}
Lisiecki, L.~E., \& Raymo, M.~E. 2005, Paleoceanography, 20,
  \dodoi{https://doi.org/10.1029/2004PA001071}

\bibitem[{{Millholland} {et~al.}(2017){Millholland}, {Wang}, \&
  {Laughlin}}]{Millholland2017}
{Millholland}, S., {Wang}, S., \& {Laughlin}, G. 2017, \apjl, 849, L33,
  \dodoi{10.3847/2041-8213/aa9714}

\bibitem[{{Mortier} \& {Collier Cameron}(2017)}]{Mortier2017}
{Mortier}, A., \& {Collier Cameron}, A. 2017, \aap, 601, A110,
  \dodoi{10.1051/0004-6361/201630201}

\bibitem[{{NASA Exoplanet Archive}(2019)}]{exoarchive}
{NASA Exoplanet Archive}. 2019, Confirmed Planets Table,  IPAC,
  \dodoi{10.26133/NEA1}

\bibitem[{{Noyes} {et~al.}(1984){Noyes}, {Hartmann}, {Baliunas}, {Duncan}, \&
  {Vaughan}}]{Noyes1984}
{Noyes}, R.~W., {Hartmann}, L.~W., {Baliunas}, S.~L., {Duncan}, D.~K., \&
  {Vaughan}, A.~H. 1984, \apj, 279, 763, \dodoi{10.1086/161945}

\bibitem[{{Pepe} {et~al.}(2013){Pepe}, {Cristiani}, {Rebolo}, {Santos},
  {Dekker}, {M{\'e}gevand}, {Zerbi}, {Cabral}, {Molaro}, {Di Marcantonio},
  {Abreu}, {Affolter}, {Aliverti}, {Allende Prieto}, {Amate}, {Avila},
  {Baldini}, {Bristow}, {Broeg}, {Cirami}, {Coelho}, {Conconi}, {Coretti},
  {Cupani}, {D'Odorico}, {De Caprio}, {Delabre}, {Dorn}, {Figueira}, {Fragoso},
  {Galeotta}, {Genolet}, {Gomes}, {Gonz{\'a}lez Hern{\'a}ndez}, {Hughes},
  {Iwert}, {Kerber}, {Landoni}, {Lizon}, {Lovis}, {Maire}, {Mannetta},
  {Martins}, {Monteiro}, {Oliveira}, {Poretti}, {Rasilla}, {Riva}, {Santana
  Tschudi}, {Santos}, {Sosnowska}, {Sousa}, {Span{\`o}}, {Tenegi}, {Toso},
  {Vanzella}, {Viel}, \& {Zapatero Osorio}}]{PePe2013}
{Pepe}, F., {Cristiani}, S., {Rebolo}, R., {et~al.} 2013, The Messenger, 153, 6

\bibitem[{{Rosenthal} {et~al.}(2021){Rosenthal}, {Fulton}, {Hirsch},
  {Isaacson}, {Howard}, {Dedrick}, {Sherstyuk}, {Blunt}, {Petigura}, {Knutson},
  {Behmard}, {Chontos}, {Crepp}, {Crossfield}, {Dalba}, {Fischer}, {Henry},
  {Kane}, {Kosiarek}, {Marcy}, {Rubenzahl}, {Weiss}, \&
  {Wright}}]{Rosenthal2021}
{Rosenthal}, L.~J., {Fulton}, B.~J., {Hirsch}, L.~A., {et~al.} 2021, arXiv
  e-prints, arXiv:2105.11583.
\newblock \doarXiv{2105.11583}

\bibitem[{{Rowe} {et~al.}(2014){Rowe}, {Bryson}, {Marcy}, {Lissauer},
  {Jontof-Hutter}, {Mullally}, {Gilliland}, {Issacson}, {Ford}, {Howell},
  {Borucki}, {Haas}, {Huber}, {Steffen}, {Thompson}, {Quintana}, {Barclay},
  {Still}, {Fortney}, {Gautier}, {Hunter}, {Caldwell}, {Ciardi}, {Devore},
  {Cochran}, {Jenkins}, {Agol}, {Carter}, \& {Geary}}]{Rowe2014}
{Rowe}, J.~F., {Bryson}, S.~T., {Marcy}, G.~W., {et~al.} 2014, \apj, 784, 45,
  \dodoi{10.1088/0004-637X/784/1/45}

\bibitem[{{Santerne} {et~al.}(2018){Santerne}, {Brugger}, {Armstrong},
  {Adibekyan}, {Lillo-Box}, {Gosselin}, {Aguichine}, {Almenara}, {Barrado},
  {Barros}, {Bayliss}, {Boisse}, {Bonomo}, {Bouchy}, {Brown}, {Deleuil},
  {Delgado Mena}, {Demangeon}, {D{\'\i}az}, {Doyle}, {Dumusque}, {Faedi},
  {Faria}, {Figueira}, {Foxell}, {Giles}, {H{\'e}brard}, {Hojjatpanah},
  {Hobson}, {Jackman}, {King}, {Kirk}, {Lam}, {Ligi}, {Lovis}, {Louden},
  {McCormac}, {Mousis}, {Neal}, {Osborn}, {Pepe}, {Pollacco}, {Santos},
  {Sousa}, {Udry}, \& {Vigan}}]{Santerne2018}
{Santerne}, A., {Brugger}, B., {Armstrong}, D.~J., {et~al.} 2018, Nature
  Astronomy, 2, 393, \dodoi{10.1038/s41550-018-0420-5}

\bibitem[{{Scargle}(1982)}]{Scargle1982}
{Scargle}, J.~D. 1982, \apj, 263, 835, \dodoi{10.1086/160554}

\bibitem[{{Schweitzer} {et~al.}(2019){Schweitzer}, {Passegger}, {Cifuentes},
  {B{\'e}jar}, {Cort{\'e}s-Contreras}, {Caballero}, {del Burgo}, {Czesla},
  {K{\"u}rster}, {Montes}, {Zapatero Osorio}, {Ribas}, {Reiners},
  {Quirrenbach}, {Amado}, {Aceituno}, {Anglada-Escud{\'e}}, {Bauer},
  {Dreizler}, {Jeffers}, {Guenther}, {Henning}, {Kaminski}, {Lafarga},
  {Marfil}, {Morales}, {Schmitt}, {Seifert}, {Solano}, {Tabernero}, \&
  {Zechmeister}}]{Schweitzer2019}
{Schweitzer}, A., {Passegger}, V.~M., {Cifuentes}, C., {et~al.} 2019, \aap,
  625, A68, \dodoi{10.1051/0004-6361/201834965}

\bibitem[{{S{\"u}veges} {et~al.}(2015){S{\"u}veges}, {Guy}, {Eyer}, {Cuypers},
  {Holl}, {Lecoeur-Ta{\"\i}bi}, {Mowlavi}, {Nienartowicz}, {Blanco},
  {Rimoldini}, \& {Ruiz}}]{Suveges2015}
{S{\"u}veges}, M., {Guy}, L.~P., {Eyer}, L., {et~al.} 2015, \mnras, 450, 2052,
  \dodoi{10.1093/mnras/stv719}

\bibitem[{{Tal-Or} {et~al.}(2019){Tal-Or}, {Trifonov}, {Zucker}, {Mazeh}, \&
  {Zechmeister}}]{Tal-Or2019}
{Tal-Or}, L., {Trifonov}, T., {Zucker}, S., {Mazeh}, T., \& {Zechmeister}, M.
  2019, \mnras, 484, L8, \dodoi{10.1093/mnrasl/sly227}

\bibitem[{{Udry} {et~al.}(2019){Udry}, {Dumusque}, {Lovis}, {S{\'e}gransan},
  {Diaz}, {Benz}, {Bouchy}, {Coffinet}, {Lo Curto}, {Mayor}, {Mordasini},
  {Motalebi}, {Pepe}, {Queloz}, {Santos}, {Wyttenbach}, {Alonso}, {Collier
  Cameron}, {Deleuil}, {Figueira}, {Gillon}, {Moutou}, {Pollacco}, \&
  {Pompei}}]{Udry2019}
{Udry}, S., {Dumusque}, X., {Lovis}, C., {et~al.} 2019, \aap, 622, A37,
  \dodoi{10.1051/0004-6361/201731173}

\bibitem[{{Valenti} \& {Fischer}(2005)}]{Valenti2005}
{Valenti}, J.~A., \& {Fischer}, D.~A. 2005, \apjs, 159, 141,
  \dodoi{10.1086/430500}

\bibitem[{{VanderPlas}(2018)}]{VanderPlas2018}
{VanderPlas}, J.~T. 2018, \apjs, 236, 16, \dodoi{10.3847/1538-4365/aab766}

\bibitem[{{Weiss} {et~al.}(2018){Weiss}, {Marcy}, {Petigura}, {Fulton},
  {Howard}, {Winn}, {Isaacson}, {Morton}, {Hirsch}, {Sinukoff}, {Cumming},
  {Hebb}, \& {Cargile}}]{Weiss2018}
{Weiss}, L.~M., {Marcy}, G.~W., {Petigura}, E.~A., {et~al.} 2018, \aj, 155, 48,
  \dodoi{10.3847/1538-3881/aa9ff6}

\bibitem[{{Zeng} {et~al.}(2019){Zeng}, {Jacobsen}, {Sasselov}, {Petaev},
  {Vanderburg}, {Lopez-Morales}, {Perez-Mercader}, {Mattsson}, {Li}, {Heising},
  {Bonomo}, {Damasso}, {Berger}, {Cao}, {Levi}, \& {Wordsworth}}]{Zeng2019}
{Zeng}, L., {Jacobsen}, S.~B., {Sasselov}, D.~D., {et~al.} 2019, Proceedings of
  the National Academy of Science, 116, 9723, \dodoi{10.1073/pnas.1812905116}

\end{thebibliography}

\appendix

\section{Statistical Properties of the Stacked Periodogram} \label{sec:appa}

{Certain properties of stacked periodograms may be derived analytically, starting with the framework for the Lomb-Scargle Periodogram \citep{Scargle1982}. These calculations provide a baseline of expectation for the strength of an aggregate signal arising from a population of low-mass planets.}

\subsection{The Lomb-Scargle Periodogram}

The Lomb-Scargle Periodogram is the following function of angular frequency $\omega$ and data $\{(t_j, X_j)\}$ for $j \in \{1,2,...,N\}$:

\begin{equation}
    P_X(\omega) = \frac{\Big[\sum_j X_j \cos \omega(t_j-\tau)\Big]^2}{2\sum_j \cos^2 \omega(t_j-\tau)} + 
                  \frac{\Big[\sum_j X_j \sin \omega(t_j-\tau)\Big]^2}{2\sum_j \sin^2 \omega(t_j-\tau)},
\end{equation}
where

\begin{equation}
    \tau = \frac{1}{2\omega}\tan^{-1}\frac{\sum_j \sin 2\omega t_j}{\sum_j \cos 2\omega t_j}.
\end{equation}
\citet{Scargle1982} demonstrated the equivalence of $2P_X(\omega)$ to

\begin{equation}
    \Delta \chi^2(\omega) = \chi^2_0 - \chi^2_m(\omega)
\end{equation}
for a sum of squares $\chi^2_0 = \sum_j X_j^2$ under a constant model, and a minimized sum of squares $\chi^2_m(\omega) = \sum_j (X_j - m(\omega))^2$ under a sinusoid model $m(\omega) = A_\omega\sin(\omega t-\phi_\omega)$ with parametrized amplitude $A_\omega$ and phase $\phi_\omega$. One may incorporate a floating mean into the sinusoid model, as well as measurement uncertainties $\sigma_j$ as weights in the summation \citep{VanderPlas2018}. Accounting for uncertainties,

\begin{equation}
    \frac{1}{2}\Delta\chi^2(\omega) = \sum_j \frac{(X_j - m(\omega))^2}{2\sigma_j^2} - \sum_j \frac{X_j^2}{2\sigma_j^2}
\end{equation}
is equivalent to the change in log-likelihood $\Delta\ln\mathcal{L}$ of residuals under a Gaussian distribution.

\subsection{Distribution of Spectral Power in a Stacked Periodogram}

Let $\hat{\mathcal{P}}(\omega)$ denote the unnormalized periodogram of $N_{\rm obs}$ RV residuals, $\{r_i\}$, as a function of angular frequency $\omega$. If the residuals are i.i.d under $r_i \sim \mathcal{N}(0, \sigma^2)$ then the periodogram values follow a $\chi^2_{\nu=2}$ distribution, or equivalently:

\begin{equation} 
\hat{\mathcal{P}}(\omega) \sim {\rm Exp}(\lambda = 1/\sigma^2)
\end{equation}
This distribution follows from the equivalence between a periodogram and a least-squares analysis of sinusoidal fits \citep{Scargle1982}. {It is important to note that the true variance $\sigma$ is unknown in practice, and usually one must use the population variance. For $N_{\rm obs} = 21$, the minimum number of RVs in any timeseries studied here, false alarm probabilities may be $\sim2\times$ higher under a more appropriate $F_{2, 20}$ distribution \citep{Koen1990}. More importantly, RV residuals are not i.i.d. in general, and the assumption of an identical, underlying variance is invalid. Therefore, the following equations only approximate the actual power in a stacked periodogram. Because RV residuals are not i.i.d. in general, it is understood that peaks in individual periodograms mandate careful treatment, such as using extreme value theory to estimate false alarm probabilities \citep{Baluev2008, Suveges2015, Delisle2020}.} 
{Next,} assume a weak signal $s(t) = A\sin({\hat{\omega} t + \phi})$ is present in an RV timeseries, $x_j = s(t_j) + r_j$. In expectation, the uncertainty-normalized sum-of-squares follows

\begin{equation} \label{eqn:sosadd}
\begin{split}
\mathbb{E} \sum_{j=1}^{N_{\rm obs}} \frac{x_j^2}{\sigma^2} & = \mathbb{E} \Big[ \sum_{j=1}^{N_{\rm obs}} \Big( \frac{r_j^2}{\sigma^2} + \frac{s(t_j)^2}{\sigma^2} + \frac{{2}s(t_j)r_j}{\sigma^2} \Big )\Big] \\ & = N_{\rm obs} + \frac{N_{\rm obs}A^2}{2\sigma^2}.
\end{split}
\end{equation}
In the limiting case that $\sigma \gg A$ the second term is negligible. The periodogram peak spans a width $\delta \omega \approx 2 \pi/T$, where $T$ is the observing baseline; although other prescriptions might be adopted for modeling peak width \citep{VanderPlas2018}. In expectation, the power input $\mathcal{S}$ from a low-amplitude sinusoid in a narrow frequency bandpass is:

\begin{equation} \label{eqn:expint}
    \begin{split}
    \mathbb{E}[\mathcal{S}] & = \mathbb{E}\Big[\int_{\hat{\omega}-\delta \omega/2}^{\hat{\omega}+\delta \omega/2} \hat{\mathcal{P}}(\omega)d\omega\Big]
    \\ & = 
    \mathbb{E}\Big[\int_{\hat{\omega}-\delta \omega/2}^{\hat{\omega}+\delta \omega/2} \frac{1}{2}(\chi^2_0 - \chi^2_s(\omega)) d\omega \Big] 
    \\& = \frac{N_{\rm obs}A^2}{4} \int_{\hat{\omega}-\delta \omega/2}^{\hat{\omega}+\delta \omega/2} {\rm sinc}^2\Big(\frac{(\omega-\hat{\omega}) \pi}{\delta\omega}\Big) d\omega \\
    & \approx \frac{N_{\rm obs}A^2}{4} \delta\omega = 2 \pi \frac{N_{\rm obs}A^2}{4T}
    \end{split}
\end{equation}
where the third equivalence invokes Fubini's theorem. The approximation is valid because $\sim 10\%$ of power lies beyond the first zero on either side of the sinc$^2(x)$ function. There is also aliasing of power by the window function $W(t)$ that lacks a simple closed form. For a stacked periodogram of $N_{\rm targets}$ (each with $N_{\rm obs}$ measurements) and the simplified case of equal measurement uncertainties, elements in the stacked periodogram follow a $\Gamma(\alpha=N_{\rm targets}, \beta=1/\sigma^2)$-distribution, neglecting statistical dependence between different frequencies from the window function. Next, define $\sigma' \equiv \sqrt{\sigma^2 + A^2/2}$ to approximate the variance introduced by a sinusoid (Equation~\ref{eqn:sosadd}). The actual density of power for a noiseless periodic signal is

\begin{equation}
    p_\omega(\mathcal{\hat{P}}; \mathcal{P}_s) = \exp{[-(\mathcal{\hat{P}}+\mathcal{P}_s)]}\sum_{m=0}^{\infty} \frac{\mathcal{\hat{P}}^{n+m-1}\mathcal{P}_s^{m}}{m!(n+m-1)!}\,,
\end{equation}
where $\mathcal{P}_s$ is the total spectral power of the signal, and $n$ is the number of frequency bins \citep{Groth1975}. However, we keep the above $\Gamma$-distribution approximation both for simplicity and because we are concerned with the regime $\sigma \gtrsim A$. If the periodic signals uniformly occupy a frequency range $\Delta \omega$, then the signal-to-noise (S/N) of a co-added periodogram point $\omega \in \Delta \omega$, relative to the remainder of the periodogram $\omega \notin \Delta \omega$, will be:

\begin{equation} \label{eqn:snr1}
    \begin{split}
    ({\rm S/N})_{\rm stack} = 2\pi\frac{\sqrt{N_{\rm targets}}N_{\rm obs}A^2}{4\sigma'^2 T \Delta \omega}.
    \end{split}
\end{equation}
A caveat to this formula is that it assumes periodogram values are statistically independent across all frequencies. In practice this assumption is untrue, since neighboring frequencies have correlated power with a lengthscale determined by the observing baseline. The correlation between distantly separated frequencies is determined by the window function \citep{Scargle1982}. {The S/N of an enhancement due to planets may be significantly reduced for complex window functions. Furthermore, the above equations assume equal measurement uncertainty, and only approximate the true distribution of power in typically heteroscedastic RV datasets.}

\end{document}